\PassOptionsToPackage{pdfpagelabels=false}{hyperref} 
\documentclass[fleqn,usenatbib]{mnras}

\usepackage{newtxtext,newtxmath}

\usepackage[T1]{fontenc}
\usepackage{ae,aecompl}

\usepackage{graphicx}	
\usepackage{amsmath}	
\usepackage{orcidlink}

\title[Characteristic time of stellar flares]{Characteristic time of stellar flares on Sun-like stars}

\author[Y. Yan et al.]{Y. Yan,$^{1,2,6,7}$\thanks{E-mail: yyan@nao.cas.cn (Y. Yan); hehan@nao.cas.cn (H. He)}
H. He$^{\orcidlink{0000-0001-9352-9189}}$,$^{1,2,8}$\footnotemark[1]
C. Li$^{\orcidlink{0000-0001-7693-4908}}$,$^{3,7}$
A. Esamdin,$^{4,8}$
B. L. Tan,$^{1,2,8}$
L. Y. Zhang$^{\orcidlink{0000-0002-2394-9521}}$$^{5}$
and H. Wang$^{6}$
\\
$^{1}$National Astronomical Observatories, Chinese Academy of Sciences, Beijing 100101, China\\
$^{2}$CAS Key Laboratory of Solar Activity, Chinese Academy of Sciences, Beijing 100101, China\\
$^{3}$School of Astronomy and Space Science, Nanjing University, Nanjing 210023, China\\
$^{4}$Xinjiang Astronomical Observatory, Chinese Academy of Sciences, Urumqi 830011, China\\
$^{5}$College of Physics, Guizhou University, Guiyang 550025, China\\
$^{6}$Center for Solar-Terrestrial Research, New Jersey Institute of Technology, University Heights, Newark, NJ 07102-1982, USA\\
$^{7}$Key Laboratory for Modern Astronomy and Astrophysics, Ministry of Education, Nanjing 210023, China\\
$^{8}$University of Chinese Academy of Sciences, Beijing 100049, China
}

\date{Accepted 2021 May 20. Received 2021 May 14; in original form 2021 April 27}

\pubyear{}

\begin{document}
\label{firstpage}
\pagerange{\pageref{firstpage}--\pageref{lastpage}}
\maketitle

\begin{abstract}
Using the short-cadence data (1-min interval) of the {\it Kepler} space telescope, we conducted a statistical analysis for the characteristic time of stellar flares on Sun-like stars (SLS). Akin to solar flares, stellar flares show rise and decay light-curve profiles, which reflect the two distinct phases (rise phase and decay phase) of the flare process. We derived characteristic times of the two phases for the stellar flares of SLS, resulting in a median rise time of about 5.9 min and a median decay time of 22.6 min. It is found that both the rise time and the decay time of the stellar flares follow a lognormal distribution. The peak positions of the lognormal distributions for flare rise time and decay time are 3.5 min and 14.8 min, respectively. These time values for stellar flares are similar to the time-scale of solar flares, which supports the idea that stellar flares and solar flares have the same physical mechanism. The statistical results obtained in this work for SLS can be a benchmark of flare characteristic times when comparing with other types of stars.
\end{abstract}



\section{Introduction}
\label{sec:intro}

The flare phenomenon on the Sun was first reported by \citet{1859MNRAS..20...13C} and \citet{1859MNRAS..20...15H}, and was described as a localized white-light flash on a time-scale of minutes. The following centuries witnessed extensive research into solar flares. A solar flare is an intense burst of radiation as a result of a sudden release of magnetic energy in a sunspot region. It is seen as a bright patch on the Sun, and can last from minutes to hours. The wavelength bands of flare radiation cover radio to $\gamma$-rays. The typical energy released in a solar flare ranges from $10^{29}$--$10^{32}$ erg \citep{2011LRSP....8....6S, 2017LRSP...14....2B}.

Analogous to solar flares, flare phenomena on other stars (namely stellar flares) have also been noticed by astronomers through inspection of stellar light curves and spectra in various bands \citep{2004A&ARv..12...71G, 2010ARA&A..48..241B}. Much bigger stellar flares, called superflares, have been observed on a variety of stars including solar-type stars \citep{1986A&A...157..217L,1989ApJ...337..927S}. \citet{2000ApJ...529.1026S} identified nine cases of superflares on normal solar-type stars (spectral class F8--G8) and reported stellar superflare energy ranging from $10^{33}$--$10^{38}$ erg. 

After decades of investigation into stellar flares, only very few flare cases have been identified for G-type stars, and even fewer for Sun-like stars (SLS; defined as stars with stellar parameters, such as effective temperature $T_{\rm eff}$ and surface gravity ${\rm log}\ g$, approximate to the Sun). Therefore, it is hard to give solid statistical results with such a small samples of flares on SLS. With the era of the {\it Kepler} space telescope \citep{2010Sci...327..977B}, many more superflares on G-type stars and SLS have been discovered by using the stellar light-curve data of {\it Kepler} \citep{2012Natur.485..478M, 2013ApJS..209....5S, 2019ApJ...871..241D, 2019ApJS..241...29Y}. It has become feasible to perform statistical analyses with a large sample of superflares on SLS. Thanks to the {\it Kepler} mission, the total monitoring time of tens of thousands of G-type and SLS objects \citep{2016RPPh...79c6901B} has become much greater than the time over which quantitative flux measurements of the Sun are available. In this way it is possible to cover the full spectrum of superflare classes and morphologies, including very large events as well as relatively rare cases, in the analyses.

Recently, the study of the solar--stellar connection has become a prominent field, developing into interdisciplinary research concerning both solar physics and stellar physics (\citealt{2003csss...12....1D, 2013PASJ...65...49S, 2017LRSP...14....4B, 2017SCPMA..60a9601C} and references therein). The long-term {\it Kepler} stellar light-curve observations greatly facilitate this field of study \citep[e.g.][]{2015ApJS..221...18H,  2016NatCo...711058K, 2017ApJ...834..207M, 2018ApJS..236....7H, 2019ApJS..244...37G, 2020Sci...368..518R, 2020ApJS..247....9Z, 2020ApJ...894L..11Z}. Among the topics, the homogeneity of solar--stellar flares is an appealing subject. Solar flares always act in a two-phase process, i.e. a rise phase and a decay phase. The rise phase generally represents a rapid release of magnetic field energy through a magnetic reconnection process, while the decay phase generally demonstrates a prolonged term comprising the whole cooling process \citep[e.g.][]{1974IAUS...57..105K, 2021MNRAS.502.3922K}. Stellar flares show similar light-curve profiles to solar flares \citep[e.g.][]{2002ApJ...580L..73G, 2021ApJ...912...81K}. In this Letter, we aim to understand the characteristic times of the two phases for stellar flares on SLS. We derive the time parameters of stellar flares by using the {\it Kepler} short-cadence (SC) data at 1-min intervals, and then conduct a statistical analysis.

The advantage of the {\it Kepler} SC data is that they can give fine profiles of stellar flares owing to the higher cadence of sampling. Thus, the time parameters of stellar flares can be derived with better accuracy and precision. \citet{2015MNRAS.447.2714B} has analysed the stellar flares observed in {\it Kepler} SC mode and found that long-duration stellar flares prefer stars with low surface gravity. \citet{2015EP&S...67...59M} studied the superflares of solar-type stars with {\it Kepler} SC data and obtained the relation between the superflare time-scale $\tau$ (defined as the e-folding decay time of flare intensity after its peak) and the flare energy $E$ (ranging from $10^{33}$--$10^{36}$ erg) as $\tau \propto E^{0.39}$. \citet{2018MNRAS.479L.139L} investigated the waiting-time (i.e. the time interval between sequential events) distribution of superflares on a solar-type star by using the {\it Kepler} SC light curves and concluded that the production of superflares is close to a stochastic Poisson process.

Utilizing the {\it Kepler} SC data and based on the solar-type (G-type) superflare star samples of \citet{2013ApJS..209....5S}, \citet{2016AcASn..57....9Y}\footnote{The paper by \citet{2016AcASn..57....9Y} is in Chinese. An English translation of the paper is available \citep{2017ChA&A..41...32Y}.} gave a preliminary result for the characteristic times of superflares on solar-type stars. Based on a relatively small sample of superflares (31 flares in total), they found that the characteristic times of the flare rise and decay phases are 0.09 and 1.0 h, respectively. In this work, we advance the analysis of stellar flare characteristic times by the following aspects: (1) expanding the number of stellar flare samples by an order of magnitude compared to \citet{2016AcASn..57....9Y}; (2) limiting the star samples to SLS for a purer distribution of flare time parameters; (3) fitting the distributions of flare time parameters with a statistical distribution function. 

This Letter is organized as follows. In Section \ref{sec:data}, we describe the observations and data reduction. In Section \ref{sec:statistics}, we present a detailed statistical analysis on the flare characteristic times of SLS. The conclusion is given in Section \ref{sec:conclusion}.

\section{Observations and data reduction}
\label{sec:data}

\subsection{Observations}

Launched in 2009, the {\it Kepler} space telescope monitored the long-term brightness variations (light curves) of the stars in a specific region of sky with an optical wavelength band of 423--897 nm \citep{2010ApJ...713L..79K}. {\it Kepler} possesses two observation modes designed for different time resolutions, i.e. the LC (long-cadence) mode with a 29.4-min cadence interval and the SC mode with a 1-min cadence interval. During the years of operation of the mission, {\it Kepler} observed more than 200000 objects in LC mode and more than 5000 objects in SC mode.

In this work we use the {\it Kepler} SC mode data, because we require high-time-resolution light curves to retrieve the time parameters of stellar flares, as explained in Section \ref{sec:intro}. The PDC (pre-search data conditioning) light-curve product of the SC mode data was used in the practical analysis, since in PDC light curves the systematic errors in the raw data have been corrected by the {\it Kepler} science pipeline.

\subsection{Data reduction}

\subsubsection{Star sample selection}

We selected SLS samples based on the flare star catalogue compiled by \citet{2015MNRAS.447.2714B}, which contains 209 stars of various types observed in {\it Kepler}'s SC mode. The SLS samples were identified from the original star catalogue by the following criteria: (1) the effective temperature is in the range of 5600 K $< T_{\rm eff} <$ 6000 K; (2) the surface gravity is in the range of $3.9 < {\rm log}\ g < 4.9$ (in cgs); (3) they should be ordinary single stars. We adopt the $T_{\rm eff}$ and ${\rm log}\ g$ values given by the {\it Kepler} Input Catalog (KIC; \citealp{2011AJ....142..112B}) for the star sample selection, as in \citet{2015MNRAS.447.2714B}. The parameter ranges of the first two criteria were chosen to place the solar values ($T_{\rm eff} \sim$ 5800 K, ${\rm log}\ g \sim$ 4.4) at the centre of the ranges, and by referring to the parameter uncertainties of the KIC ($\pm 200$ K for $T_{\rm eff}$, $\pm 0.5$ dex for ${\rm log}\ g$). For the third criterion, we excluded objects labelled `binary' in the catalogue of \citet{2015MNRAS.447.2714B} and in the {\it Kepler} archive.

Based on the above criteria, we finally picked out 20 SLS samples with unambiguous flare peaks in their light curves. The full list of the selected SLS samples as well as their stellar parameters is given in the online supplementary material (`Table of star samples'). All of the selected SLS objects are relatively bright stars with {\it Kepler} magnitudes in the range of [7.397, 11.662] (see the `KeplerMag' column of the SLS sample table in the online supplementary material). The per-point measurement errors of SC data for {\it Kepler} objects in this magnitude range are equal to or better than 200 parts per million (ppm) per min \citep{2010ApJ...713L.160G}.

\subsubsection{Flare sample identification}

We identified stellar flares by visual inspection on the SC-PDC light curves of the 20 selected SLS samples. In the first run of recognition, we got approximately 200 preliminary flare samples. However, some of them have issues of low signal-noise ratios, overlapping of flare signals, or data gaps, and are not appropriate for the analysis of flare time properties. After removing the inappropriate samples, 184 flares remained, which were utilized for the subsequent analysis.

\begin{figure}
 \includegraphics[width=\columnwidth]{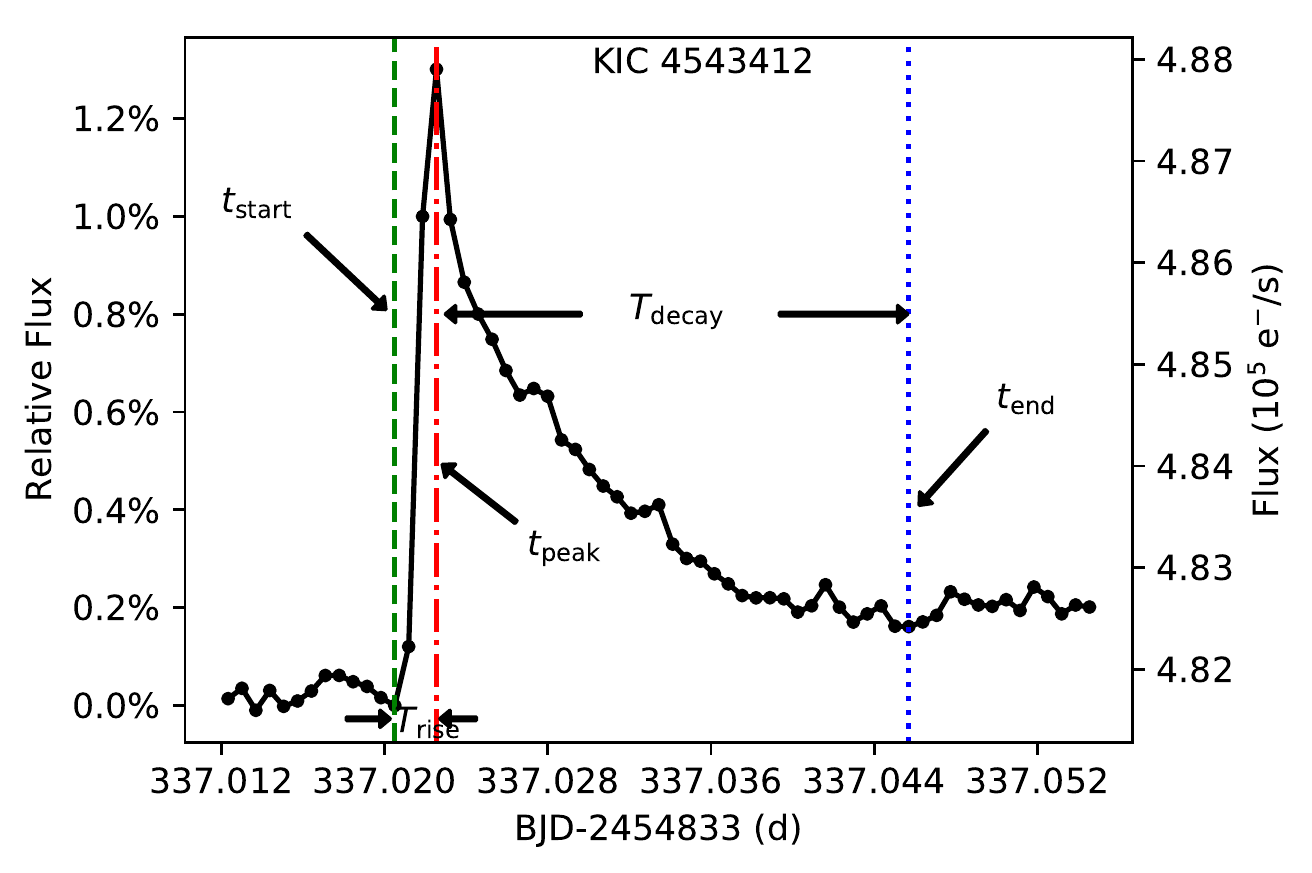}
 \centering
 \caption{An example flare light curve of an SLS (KIC 4543412) observed in {\it Kepler} SC mode. The dots on the curve represent the SC data points at 1-min intervals. The time parameters ($t_{\rm start}$, $t_{\rm peak}$, $t_{\rm end}$, $T_{\rm rise}$, and $T_{\rm decay}$) of the flare are marked in the plot. The percentage value of the flare intensity enhancement (relative flux) shown on the left {\it y}-axis is relative to the absolute flux value (shown on the right {\it y}-axis) at the time of $t_{\rm start}$. BJD in the {\it x}-axis means barycentric Julian date. The offset 2454833 is the Julian date on 2009 January 1.}
 \label{fig:para}
\end{figure}

An example flare light curve from the identified flare samples of SLS is shown in Fig. \ref{fig:para}. The full list of 184 flares and all the light-curve plots of the flare samples are given in the online supplementary material (`Table of flare samples' and `Plots of all the flare samples', respectively). The numbers of flare samples for individual SLS are included in the SLS sample table in the online supplementary material. The energy range of the flare samples is $10^{33}$--$10^{36}$ erg \citep{2015MNRAS.447.2714B}.

\subsubsection{Deriving the rise time and decay time of each flare}

The time parameters of the identified flare samples were derived based on the flare light-curve profiles. In this work, we focus on the durations of the flare rise phase and decay phase. We first determined the start, peak, and end times (denoted as $t_{\rm start}$, $t_{\rm peak}$, and $t_{\rm end}$) of each flare, as demonstrated in Fig. \ref{fig:para}, by visually inspecting the flare light curves. The definitions of these times are as follows: $t_{\rm start}$ is the time of the data point where the flare flux suddenly rises from the background level; $t_{\rm peak}$ is the time of the data point at which the flare intensity reaches its peak; and $t_{\rm end}$ is the time of the data point where the flare flux is asymptotically tangent to the background level. Then, the flare rise time and decay time (denoted as $T_{\rm rise}$ and $T_{\rm decay}$) can be derived by $T_{\rm rise} = t_{\rm peak} - t_{\rm start}$ and $T_{\rm decay} = t_{\rm end} - t_{\rm peak}$, respectively (see illustration in Fig. \ref{fig:para}). The uncertainty of the derived $T_{\rm rise}$ and $T_{\rm decay}$ values is 1-min due to the cadence interval of the SC data. Note that, when deriving the flare time parameters, we have no presupposition of specific shapes for the flare rise phase or decay phase.

The obtained time parameters of all the flare samples are included in the flare sample table and illustrated in the flare light-curve plots in the online supplementary material. Because the accurate cadence-interval value of the {\it Kepler} SC mode is 58.86 s \citep{2010ApJ...713L..79K}, the derived values of $T_{\rm rise}$ and $T_{\rm decay}$ shown in the flare list table may not be exact multiples of 1 min.

\section{Statistical analysis of the flare characteristic times of SLS}
\label{sec:statistics}

\subsection{Evaluating statistical quantities for the flare rise time and decay time}
\label{sec:stat-quantities}

We performed a statistical analysis on the two time parameters $T_{\rm rise}$ and $T_{\rm decay}$ for the 184 flare samples of SLS. Table \ref{tab:stat} gives the results of the general statistical quantities, including median, mean, first quartile (Q1)\footnote{The first quartile (Q1) is the median of the lower half of a data set.}, third quartile (Q3)\footnote{The third quartile (Q3) is the median of the upper half of a data set.}, minimum (Min), maximum (Max), and standard deviation (Std) values. 

\begin{table}
\caption{Statistics for the rise time and decay time of the flare samples of SLS}
\label{tab:stat}
\begin{tabular}{c c c c c c c c}
\hline
 & Median & Mean & Q1 & Q3 & Min & Max & Std \\
\hline
$T_{\rm rise}$  & 5.9 & 8.8 & 3.9 & 9.1 & 1.0 & 152.0 & 13.2 \\
$T_{\rm decay}$  & 22.6 & 33.7 & 15.7 & 37.3 & 4.9 & 216.8 & 32.4 \\
\hline
\end{tabular}
\\
Note: See Section \ref{sec:stat-quantities} for the meanings of the quantities. All of the values are in minutes.
\end{table}

Most flares of SLS have a small value of $T_{\rm rise}$, with a minimum of 1.0 min and a median of 5.9 min; about 25\% of the $T_{\rm rise}$ values are within 3.9 min (see Q1 column in Table \ref{tab:stat}), and about 75\% of the $T_{\rm rise}$ values are within 9.1 min (see Q3 column in Table \ref{tab:stat}). The small differences between median versus Q1 and Q3 versus Q1 suggest that the values of $T_{\rm rise}$ are concentrated at the head. Therefore, flares of SLS are quite similar to solar flares in this tendency, i.e. most flares have an abrupt and rapid rise phase. However, there are still some longer ones with a maximum of 152.0 min, though at a low occurrence rate. Flare events with long rise times have also been observed on the Sun \citep[e.g.][]{1999ApJ...519L..93M}, but as in the case of stellar flares, these events are not very frequent.

The $T_{\rm decay}$ values of the flares of SLS also behave like those seen in solar flares. They spread over a range of 4.9--216.8 min with a median of 22.6 min. The larger median value of $T_{\rm decay}$ indicates that flares tend to have a more gradual decay phase than rise phase. The Q1 and Q3 quantities of $T_{\rm decay}$ suggest the same tendency as $T_{\rm rise}$; that is, the values of $T_{\rm decay}$ are also concentrated at the head with longer ones in the tail.

\subsection{Fitting the distributions of flare rise time and decay time}
\label{sec:fitting}

\begin{figure*}
 \includegraphics[width=0.72\textwidth]{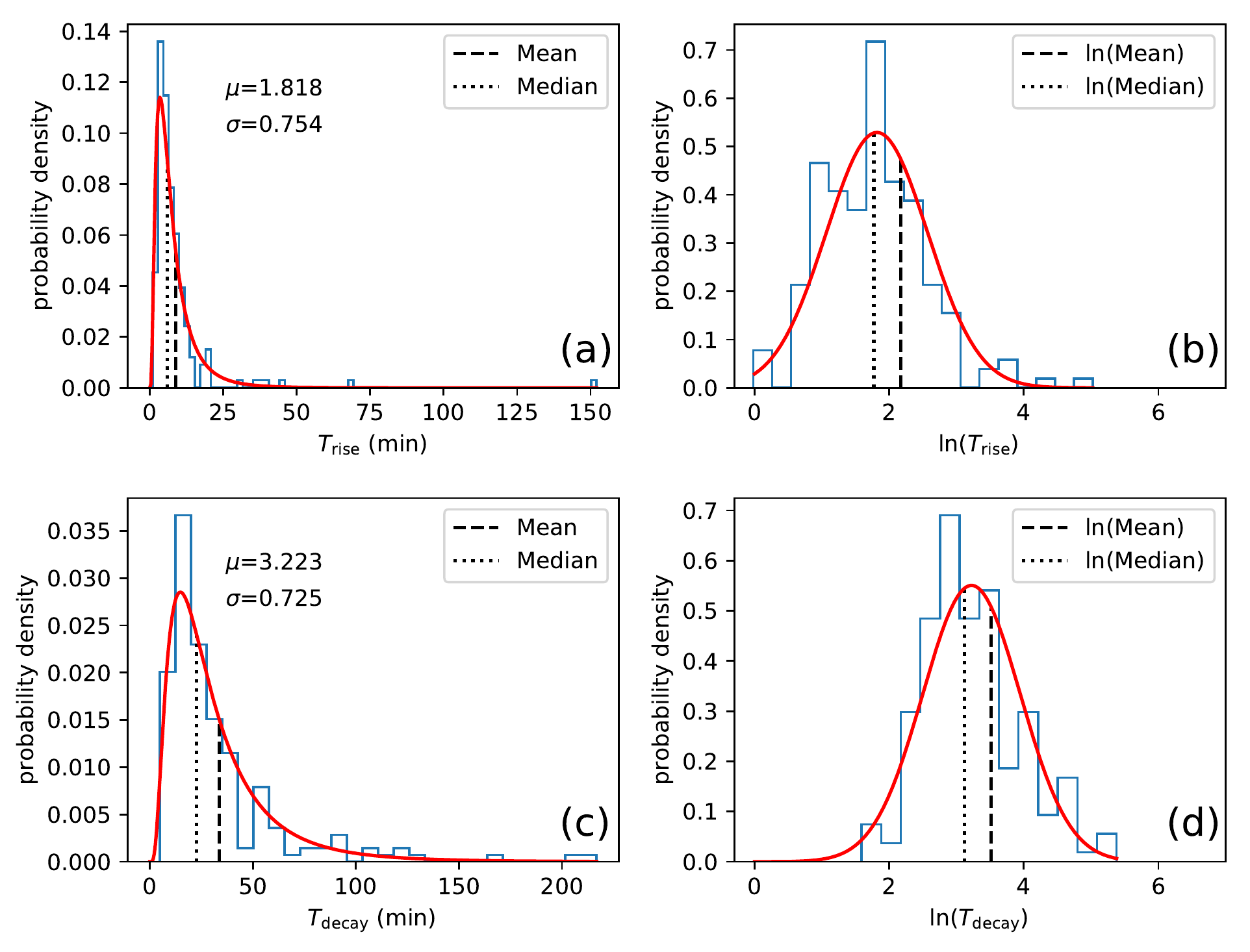}
 \caption{(a), (c) Histograms (blue) and fitted log-normal distribution curves (red) for $T_{\rm rise}$ and $T_{\rm decay}$. The vertical dashed / dotted lines indicate the statistical mean / median values of the $T_{\rm rise}$ and $T_{\rm decay}$ data sets, as listed in Table \ref{tab:stat}. $\mu$ and $\sigma$ are the two parameters of the fitted log-normal distribution, as listed in Table \ref{tab:fit}. (b), (d) Histograms (blue) and fitted normal distribution curves (red) for $\ln(T_{\rm rise})$ and $\ln (T_{\rm decay})$. The vertical dashed / dotted lines indicate the statistical $\ln ({\rm mean})$ / $\ln ({\rm median})$ values of the $T_{\rm rise}$ and $T_{\rm decay}$ data sets, which are not necessarily an exact match to the fitted curves.}
 \label{fig:fitting}
\end{figure*}

One can see that both $T_{\rm rise}$ and $T_{\rm decay}$ have a noticeable variance between the values of median and mean shown in Table \ref{tab:stat}, which suggests that there is a deviation from the normal distribution. In the left-hand panels of Fig. \ref{fig:fitting}, we display the distributions of $T_{\rm rise}$ (a) and $T_{\rm decay}$ (c) with histograms. These plots confirm the results obtained by the statistical quantities in Section \ref{sec:stat-quantities} that the distributions of both $T_{\rm rise}$ and $T_{\rm decay}$ have peak-shaped heads and long tails, which is consistent with the morphological properties of the log-normal distribution. To verify this impression, we show the distribution diagrams of $\ln (T_{\rm rise})$ and $\ln (T_{\rm decay})$ in Figs \ref{fig:fitting}(b) and (d). If $T_{\rm rise}$ and $T_{\rm decay}$ follow a lognormal distribution, then $\ln (T_{\rm rise})$ and $\ln (T_{\rm decay})$ should take the form of a normal distribution. The morphologies of the $\ln (T_{\rm rise})$ and $\ln (T_{\rm decay})$ histograms in Fig. \ref{fig:fitting} affirm this idea.

We fitted the distributions of $T_{\rm rise}$ and $T_{\rm decay}$ with the log-normal probability density function (PDF), which can be expressed as
\begin{equation} \label{equ:lognormal}
    P(x) = \frac{1}{x\sigma\sqrt{2\pi}} {\rm e}^{-\frac{(\ln x - \mu)^2}{2\sigma^2}}.
\end{equation}
The fitted log-normal PDFs for $T_{\rm rise}$ and $T_{\rm decay}$ are displayed with red curves in Figs \ref{fig:fitting}(a) and (c). We also fitted the normal distributions for $\ln (T_{\rm rise})$ and $\ln (T_{\rm decay})$, and the resultant curves are displayed in Figs \ref{fig:fitting}(b) and (d). 

As shown by equation (\ref{equ:lognormal}), a log-normal PDF can be characterized by two parameters, $\mu$ and $\sigma$, in which $\mu$ defines the natural logarithm of the median value of the distribution, and $\sigma$ defines the shape of the distribution. The $\mu$ and $\sigma$ values of the fitted log-normal PDFs for $T_{\rm rise}$ and $T_{\rm decay}$ are given in Table \ref{tab:fit}. The median (${\rm e}^{\mu}$) and mode (${\rm e}^{\mu-\sigma^2}$; indicating peak position) values of the fitted lognormal distributions are also listed in Table \ref{tab:fit}.

\begin{table}
\caption{Lognormal fitting parameters for the rise time and decay time of the flare samples of SLS}
\label{tab:fit}
\begin{tabular}{c c c c c c c}
\hline
 & $\mu$ & $\sigma$ & ${\rm e}^{\mu}$ & ${\rm e}^{\mu-\sigma^2}$ & $D$ & $D_{\rm N, 0.05}$  \\
\hline
$T_{\rm rise}$  & 1.818 & 0.754 & 6.2 & 3.5 & 0.0819 & 0.1003 \\
$T_{\rm decay}$  & 3.223 & 0.725 & 25.1 & 14.8 & 0.0835 & 0.1003 \\
\hline
\end{tabular}
\\
Note: See Sections \ref{sec:fitting} and \ref{sec:K-S} for the meanings of the quantities. ${\rm e}^{\mu}$ and ${\rm e}^{\mu-\sigma^2}$ (median and mode values of the fitted log-normal distribution) are in minutes.
\end{table}

\subsection{Kolmogorov--Smirnov test for the fitted distributions}
\label{sec:K-S}

The Kolmogorov--Smirnov (K--S) test is a useful tool to evaluate the accuracy with which the data follow the fitted distribution. For the data sets of $T_{\rm rise}$ and $T_{\rm decay}$, we first calculated the quantity statistic $D$, which is the maximum distance between the cumulative density function (CDF) of the fitted distribution and the CDF of the data distribution, and then compared $D$ with critical values $D_{N,\alpha}$ ({\it N}: count of samples; $\alpha$: significance level) of the K--S test.

The $D$ values that we obtained for $T_{\rm rise}$ and $T_{\rm decay}$ are 0.0819 and 0.0835, respectively. According to the critical value table of the K--S test, for a large sample number and with $\alpha = 0.05$, the $D_{N,\alpha}$ value can be calculated by 
\begin{equation}
    D_{N, 0.05} = \frac{1.36}{\sqrt{N}}.
\end{equation}
In this work, the sample size is 184, so that $D_{N, 0.05} = 0.1003$. All the values of $D$ and $D_{N, 0.05}$ for $T_{\rm rise}$ and $T_{\rm decay}$ are tabulated in the final two columns of Table \ref{tab:fit}. It is seen that for both $T_{\rm rise}$ and $T_{\rm decay}$, $D < D_{N, 0.05}$. Thus, we can say that both the data sets of $T_{\rm rise}$ and $T_{\rm decay}$ follow a lognormal distribution at a significance level of 0.05.

\section{Conclusion}
\label{sec:conclusion}

Using the {\it Kepler} SC light-curve data with a 1-min cadence interval, we conducted a statistical analysis for the characteristic time of stellar flares on SLS. Because the rise phase and decay phase of a flare correspond to two distinct physical processes (a rapid energy release process versus a cooling process), we analyze the characteristic times of the two phases separately. We identified 184 stellar flares from the SC light curves of 20 SLS samples. We derived the rise time, $T_{\rm rise}$, and decay time, $T_{\rm decay}$, of each flare, and performed statistical analyses on the two time parameters. The conclusions are as follows:

(a) The median values of $T_{\rm rise}$ and $T_{\rm decay}$ are 5.9 and 22.6 min, respectively, and the mean values of $T_{\rm rise}$ and $T_{\rm decay}$ are 8.8 and 33.7 min, respectively (see Table \ref{tab:stat}). The ratio of the time-scale (reflected by the median or mean values) of the flare rise phase to that of the decay phase (denoted as $\tau_{\rm rise}:\tau_{\rm decay}$) is roughly 1:4. In addition, the ratio of the first quartile (Q1) values and the ratio of the third quartile (Q3) values between $T_{\rm rise}$ and $T_{\rm decay}$ are also roughly 1:4 (see Table \ref{tab:stat}). This result for the characteristic times of stellar flares on SLS updates the values in \citet{2016AcASn..57....9Y} with a much larger data set of flare samples.

(b) Both $T_{\rm rise}$ and $T_{\rm decay}$ follow a lognormal distribution, showing a peak-shaped head and a long tail (see Fig. \ref{fig:fitting}), which can be characterized by two log-normal parameters, $\mu$ and $\sigma$ (see Table \ref{tab:fit}). The peak positions (i.e. mode values ${\rm e}^{\mu-\sigma^2}$) of the fitted log-normal distributions for $T_{\rm rise}$ and $T_{\rm decay}$ are 3.5 and 14.8 min, respectively (see Table \ref{tab:fit}). The results are independent of specific shapes of flare rise phase or decay phase.

(c) \citet{2001A&A...375.1049T} analysed the characteristic time of solar flares observed in the ${\rm H}\alpha$ band (a narrow band centred at 656.3 nm; solar flares observed in this band also called optical flares and are comparable with the stellar flares observed in the {\it Kepler} band). They found that for the solar flare samples with `importance class 1', the median values of the flare rise time and decay time are 5.0 and 22.0 min, respectively. These characteristic times of solar flares are similar to the time values obtained in this work for stellar flares of SLS, even though the energy of the stellar flares ($10^{33}$--$10^{36}$ erg) is much greater than that of solar flares ($10^{29}$--$10^{32}$ erg). This supports the idea that stellar flares and solar flares have the same physical mechanism.

The statistical results obtained in this work for SLS can be a benchmark of flare characteristic times when compared with other types of stars. Besides the data of {\it Kepler}, stellar light curves obtained by the {\it TESS} space telescope \citep{2015JATIS...1a4003R} and the future {\it PLATO} mission \citep{2014ExA....38..249R} can be utilized for this sort of analysis. The observations by ground-based telescopes also play an important role in stellar flare study \citep[e.g.][]{2021RAA....21....7B}. Stellar flare radiation is a key factor in the habitability of exoplanets within a stellar system \citep{2020IJAsB..19..136A}. The analysis results of stellar flare 
activity properties will be useful for this topic of research.

\section*{Acknowledgements}
We are grateful to the reviewer for their valuable comments that improved the paper. This work is supported by the National Natural Science Foundation of China (11973059) and the National Key R\&D Program of China (2019YFA0405000). YY is grateful for the support of the visiting scholar programme provided by CAS. YY, HH, and CL are grateful to the open research programme entitled ``The statistical analysis on white-light flares from solar-like stars'', funded by the Key Laboratory of Modern Astronomy and Astrophysics (Nanjing University), Ministry of Education. HH acknowledges the B-type Strategic Priority Program of the Chinese Academy of Sciences (XDB41000000), the CAS Strategic Pioneer Program on Space Science (XDA15052200), and the Astronomical Big Data Joint Research Center, co-founded by the National Astronomical Observatories, Chinese Academy of Sciences and the Alibaba Cloud. CL thanks the National Natural Science Foundation of China (grant 11673012) and the ``Integration of Space- and Ground-based Instruments'' project of the China National Space Administration. LYZ thanks the National Natural Science Foundation of China (grant 11693002). \textsc{python}, \textsc{matplotlib}, \textsc{numpy}, \textsc{astropy}, \textsc{scipy}, and \textsc{pandas} were used in this study. This paper includes data collected by the {\it Kepler} mission and obtained from the Mikulski Archive for Space Telescopes (MAST) at the Space Telescope Science Institute (STScI). Funding for the {\it Kepler} mission is provided by the NASA Science Mission Directorate. STScI is operated by the Association of Universities for Research in Astronomy, Inc., under NASA contract NAS 5–26555.

\section*{Data Availability}

{\it Kepler} SC data can be accessed from MAST via \url{https://doi.org/10.17909/T90K5C}. The data underlying this article are available in the online supplementary material.


\bibliographystyle{mnras}
\bibliography{reference} 



\bsp	
\label{lastpage}
\end{document}